\documentstyle[aps,epsfig,twocolumn]{revtex}

\begin{document}
\newcommand\rootiii{($\sqrt{3}\!\times\!\sqrt{3})$}
\newcommand\iiixi{(3$\times$1)}
\newcommand\vixi{(6$\times$1)}
\newcommand\ixi{(1$\times$1)}
\newcommand\sbs{(7$\times$7)}

\title{Ag induced zero- and one-dimensional nanostructures on vicinal Si(111)}

\author{J. Kuntze\cite{correspondingauthor}, A. Mugarza, J. E. Ortega}
\address{Departamento de F\'{\i}sica
Aplicada I, Universidad del Pa\'{\i}s Vasco, Plaza de O\~nate 2,
E-20018 San Sebastian, Spain}
\address{Donostia International Physics Center and Centro Mixto
CSIC/UPV, Paseo Manuel Lardizabal 4, 20018 San Sebastian, Spain}

\date{\today}
\maketitle

\begin{abstract} The formation of a Ag stabilized
regular step lattice on vicinal Si(111) miscut towards
[11$\bar{2}$] is reported. The step bunching characteristic of the
clean surface is prevented by a single-domain Si(111)-\iiixi-Ag
reconstruction. The nanostructured surface is used as a template
for growing one-dimensional arrays of 1 nm sized Ag quantum dots
with a preferential spacing of 1.5 nm along the rows.\\ \noindent
PACS number(s): 68.37.Ef, 81.16.Dn, 81.07.Ta, 81.07.Vb
\end{abstract}
\vspace{1ex}

Self-organization on crystal surfaces is a promising alternative
for growing uniform nanostructures and templates with regular
sizes and spacings \cite{shchukin_99a}. In particular vicinal
surfaces appear as natural substrates for the self-assembly of
linear structures. Indeed, arrays of quantum wires and dots can be
obtained by step decoration and step-flow growth
\cite{himpsel_99a,gambardella_02a}. Also one-dimensional templates
can be tailored at vicinal surfaces by inducing periodic faceting
with adsorbates \cite{bachmann_01a,hild_02a} or simply by
stabilizing single domains of atomic row structures at terraces
\cite{himpsel_99a,bennewitz_02a}. The latter is the case of
self-assembled (5$\times$2) Au rows on vicinal Si(111) with small
(1$^{\circ}$) miscut, which can serve as read-writing tracks for
atomic scale memories \cite{bennewitz_02a}.

Here we investigate how surface reconstruction and morphology of
Si(111) vicinals miscut towards [11$\bar{2}$] are affected by Ag
adsorption with special emphasis on nanostructure and device
fabrication. We find that the \iiixi-Ag reconstruction prevents
the equilibrium step bunching characteristic of the clean surface
at 870 K. The system displays a regular array of monatomic steps
with terraces that contain three equidistant Ag atomic chains.
Further deposition of Ag at RT results in linear arrays of quantum
dots nucleating in the trenches between the Ag rows.

The experiments have been performed in an UHV system (base
pressure below 5$\times$10$^{-11}$ mbar) equipped with a
commercial Omicron scanning tunneling microscope (STM) and
low-energy electron diffraction (LEED). Samples were cut from
n-type Si(111) wafers (0.05--0.1 $\Omega$cm) with a 4$^\circ$ and
6.1$^\circ$ miscut towards [11$\bar{2}$]. The clean surface was
prepared by direct current heating with the current flowing
parallel to the step direction to prevent electromigration effects
on the step distribution \cite{latyshev_89a,yagi_01a}. Surface
temperature was monitored by an infrared pyrometer with the
emissivity set to 0.4. The wafers were initially outgassed at 970
K for 1 h, followed by repeated flashing to 1500 K for a few
seconds to remove the oxide layer and SiC. To produce ordered
arrays of step bunches, we followed a procedure similar to the one
given by Lin {\it et al.}\cite{lin_98a}: flashing to 1500 K (10
s), followed by a fast (10 s) cooling to 1140 K (slightly above
the (1$\times$1) to (7$\times7$) transition for Si(111) vicinal
towards [11$\bar{2}$]\cite{phaneuf_88a}) and a slow cooling ($\le$
1 K/s) to 920 K, where the sample was held for a further
postanneal of 30 min to allow step bunches to order. The pressure
during the whole procedure never exceeded 3$\times$10$^{-10}$
mbar. Ag was evaporated from an e-beam source with the deposition
being monitored by a flux-meter calibrated against a crystal
balance. Typical evaporation rates were 0.3 ML/min. The pressure
during evaporation was in the 10$^{-11}$ mbar range.

\begin{figure}
\centerline{\epsfig{figure=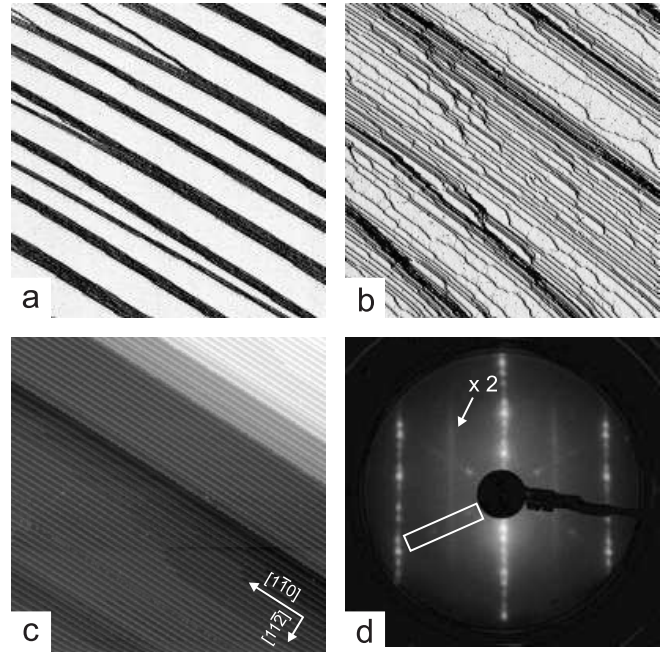,width=\linewidth}}
\vspace{1ex} \caption{\label{fig1} (a) STM image of the clean
surface. The image is differentiated to enhance the step bunches.
Step-down direction is from lower left to upper right, size 600
nm. (b) (partially) debunched surface after Ag adsorption and
redesorption (cf. text). Differentiation and step-down direction
 as in (a), size 350 nm. (c) Regular step array after \iiixi\
formation. Each line corresponds to a single step edge [cf. Fig.
\ref{fig2}(b)]. Occasional multiple steps separate regular stepped
regions. Step-down direction is along [11$\bar{2}$], size 200 nm.
(d) LEED pattern of the \iiixi-reconstructed surface at 26 eV.
Faint residues of other rotational \iiixi\ domains are indicated
by a box. The weak vertical stripes (arrow) are due to buckling of
step edge atoms [cf. text and Fig. \ref{fig2}(b)].}
\end{figure}

Vicinal Si(111) surfaces miscut towards [11$\bar{2}$] undergo
spontaneous step bunching \cite{phaneuf_88a,phaneuf_87a}. Figure
\ref{fig1}(a) shows a large scale image of the clean surface with
an array of bunched steps. The total periodicity of stepped
bunches and flat (111) terraces reaches a self-limited size of
approximately 70$\pm$10 nm in agreement with previous reports
\cite{lin_98a,phaneuf_91a,men_02a}.

Deposition of submonolayer coverages of Ag and subsequent
annealing to 850-900 K leads to partial erosion of the ordered
step arrays, triggered by the formation of \iiixi\ domains. The
\iiixi\ structure is oriented almost exclusively along the step
direction, contrary to Ag induced structures on flat Si(111),
where all three domains of \iiixi\ are observed, usually in
addition to coexisting \rootiii\ and \sbs\ phases
\cite{wan_93a,hasegawa_99a,sakamoto_02a}.

To optimize the long-range order and to maximize the \iiixi\
coverage in one domain, we investigated the elementary preparation
steps more closely. Figure \ref{fig1}(b) shows the same surface as
(a) after subsequent deposition of 1 ML Ag followed by annealing
to 870 K for 30 min. The Ag has desorbed and the surface has
transformed to (7$\times$7) reconstructed terraces of varying
width, separated by monatomic steps which have partially coalesced
to poorly defined bunches. We conclude that Ag promotes debunching
of the surface upon annealing, since no debunching is observed
directly after Ag deposition at RT. Since annealing above the Ag
desorption temperature of 820 K \cite{wan_93a} is needed for
sufficient step mobility, the surface morphology is kinetically
determined by the competing factors of Ag promoted debunching and
(re-)bunching of the clean terraces after Ag desorption. Thus, in
this case good long-range ordering cannot be achieved by mere Ag
deposition and annealing.

To facilitate better ordering, Ag was deposited on the clean
surface after rapid ($\approx$100 K/s) quenching to RT from above
the \ixi\ to \sbs\ transition temperature, minimizing the number
and size of preexisting step bunches. STM images of the quenched
surface prior to Ag deposition (not shown) reveal a surface
morphology similar to the one in Fig. \ref{fig1}(b).

After deposition of approximately 0.3 ML Ag and subsequent
annealing to 870 K for some 10 s, the \iiixi\ reconstruction is
formed. Cycling this desorption/annealing sequence a few times
further improves long-range order, overcoming the limitations of
mass transport at this temperature. Alternatively, Ag can also be
deposited at 870 K, but special care has to be taken to balance
deposition and desorption rates. Figure \ref{fig1}(c) demonstrates
the surface quality achieved by this preparation. On the
mesoscopic ($\mu$m$^2$) scale the main part ($>$90\%) of the
surface is covered by only one \iiixi\ domain. Only occasionally
other rotational domains are found extending over only small
areas. Due to the size of the electron beam in our LEED setup (2
mm), faint residues of such rotational domains are usually found
in LEED images, as demonstrated in Fig. \ref{fig1}(d).
Additionally, near the cathode and anode edges of the sample, we
find residual \rootiii\ or \sbs\ domains, respectively. We
attribute this to electromigration of Ag towards the cathode
\cite{shi_00a}, since the observed distribution depends on the
current direction. Heating by electron bombardement should
eliminate this migration effect.

The \iiixi\ spots in the LEED image [Fig. \ref{fig1}(d)] are split
due to superlattice diffraction \cite{henzler_76a}. From the
splitting an average step separation of 4.2$\pm$0.3 nm is derived
\cite{leedremark}, compatible with the dominant terrace width of
approximately 3.8 nm observed by STM in Fig. \ref{fig1}(c).
Incorporation of wider or narrower terraces, varying in width by
multiples of the \iiixi\ unit cell size, accounts for the misfit
of the local slope (4.7$^\circ$ for 3.8 nm terraces separated by
single bilayer steps) to the macroscopic miscut. This
 leads to residual imperfections of the step lattice over large
($\mu$m$^2$) areas, which could eventually be overcome by matching
the macroscopic miscut to preferred terrace sizes.

\begin{figure}
\centerline{\epsfig{figure=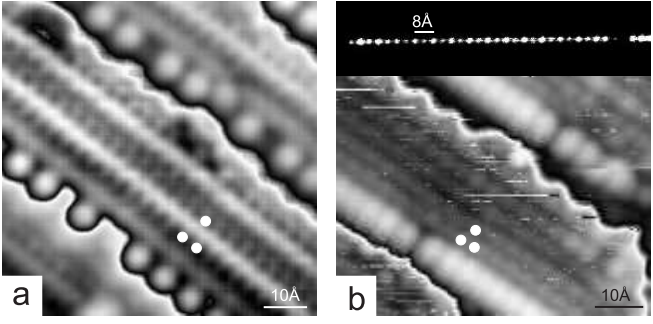,width=\linewidth}}
\vspace{1ex} \caption{\label{fig2} High-resolution STM images of a
few \iiixi\ reconstructed terraces. Each terrace is shown in a
full grayscale for clarity. Step-down direction is from upper
right to lower left. Three marks in each image indicate the
position of Si related maxima in calculated STM images
\protect\cite{erwin_98a}. (a) empty state image ({\it
U$_{tip}$}=-1 V, {\it I}=50 pA, size 7.7 nm). The mark "A"
indicates Si adatoms. The bright rows on the terrace correspond to
the Ag atom sites. (b) filled state image ({\it U$_{tip}$}=1.5 V,
{\it I}=100 pA, size 5nm). The bright step edge row is probably
due to Si states. It exhibits a slight buckling of 0.05 \AA (see
inset, where a single row is shown with better contrast).}
\end{figure}

Another weak feature in the LEED image of Fig. \ref{fig1}(d) are
faint streaks at $\times$2 position along the [11$\bar{2}$]
azimuth, i.e., parallel to the step edge direction. These can be
explained by high-resolution STM images: Figure \ref{fig2}
displays a close-up view of a few \iiixi\ reconstructed terraces
for both filled and empty states. In the empty states image [Fig.
\ref{fig2}(a)] ball-like features (mark "A") are located at step
edges in the step down-direction. These are probably due to Si
adatoms with a two nearest neighbor spacing along the step edge
direction ([$\bar{1}$10]). The $\times$2 spacing is likely the
cause for a buckling of step edge atoms with the same periodicity,
which show up as prominent bright rows in filled state images
[Fig. \ref{fig2}(b)]. In the inset a single step edge row is seen
with higher contrast. The difference in apparent height of
alternating bright and shallow atoms along the row is 0.05 \AA. We
propose that the bright step edge rows and the adatom features
both correspond to Si atoms rather than Ag, since upon \iiixi\
formation excess Si atoms from the \sbs-phase have to be
incorporated at steps \cite{saranin_99a}. The strong bias
dependence of the adatoms and the nucleation of Ag on the step
edge rows (see below)  further support this assignment.

The marked triangles in both filled and empty states images
correspond to maxima due to Si surface states in calculated STM
images of the metal induced \iiixi\ reconstruction in the
honeycomb chain-channel (HCC) model proposed by Erwin and
Weitering \cite{erwin_98a}. The additional bright row seen in
empty states images [Fig. \ref{fig2}(a)] corresponds to the metal
atom position and is due to a superposition of states from Si and
metal atoms. We note that Erwin and Weitering have found a
distortion of the Si honeycombs for the case of Ag leading to a
\vixi\ symmetry in agreement with previous
\cite{wan_93a,sakamoto_02a,saranin_99a,carpinelli_95a} works. In
STM images, the nearly equilateral triangle of filled state maxima
[cf. marks in Fig. \ref{fig2}(b)] was found to be distorted by
pairing of maxima in neighboring rows. The direction of this
pairing alternates in neighboring \iiixi\ unit cells, leading to
the \vixi\ periodicity. For the stepped surface, we observe the
same distortion on occasional wider terraces. On the average
terrace of approximately 3.8 nm width, we do not observe \vixi\
formation. The small terrace size could facilitate strain relief
by relaxation near steps, thus eliminating the driving force for
\vixi\ formation \cite{erwin_98a}.

\begin{figure}
\centerline{\epsfig{figure=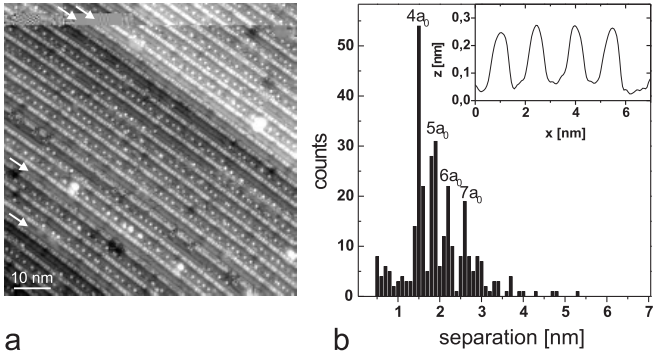,width=\linewidth}}
\vspace{1ex} \caption{\label{fig3} (a) Array of Ag nanodots grown
on a \iiixi-stabilized step lattice [cf. Fig.
\protect\ref{fig1}(c)] after RT deposition of approx. 0.5 ML Ag.
The arrows mark terraces narrower than 3.8 nm where no dots
nucleate. The distribution of the dot spacings is shown in (b).
Pronounced peaks indicate preferred separations along
[$\bar{1}$10], with a clear maximum at 4a$_0$ (a$_0$=0.384 nm). A
section across four dots is seen in the inset, displaying the
uniform shape of approximately 1 nm diameter and 0.2 nm height.}
\end{figure}

After fabrication and characterization of the stepped
Ag/Si(111)-\iiixi\ surface, we can use this array as a template
for further Ag growth. As shown recently for flat Si(111),
one-dimensional arrays of Ag nanodots nucleate at RT on \iiixi-Ag
reconstructed areas with a preferential spacing of 3a$_0$--5a$_0$,
when a$_0$=0.384 nm is the lattice spacing along $<110>$
\cite{hirayama_01a}. We find a very similar behaviour on stepped
templates (Fig. \ref{fig3}), with uniform Ag nanodots of 1 nm
diameter and 0.2 nm height nucleating in the trenches between the
Ag rows in the \iiixi\ structure. The nearest neighbor distance is
preferentially 4a$_0$, with multiples up to 7a$_0$ being observed.
The existence of clear peaks in the distribution [Fig.
\ref{fig3}(b)] suggests a nucleation at well defined lattice
sites. Strictly one-dimensional lines of dots are found on 3.8 nm
wide terraces, whereas narrower terraces prevent nucleation
[arrows in Fig. \ref{fig3}(a)]. On wider terraces (not shown), a
second line of Ag dots grows parallel to the first. The structural
characteristics can thus be fine tuned by controlling the terrace
size, i.e., by different macroscopic miscuts.

In addition to nanodot nucleation on the terraces, we observe
nucleation on the bright step edge row of the \iiixi\ structure
[cf. Fig. \ref{fig2}(b)]. The nucleated rows again are strictly
one-dimensional, but the dots display a less regular shape and are
less well ordered along [$\bar{1}$10]. The Ag nucleation at this
sites provides further evidence that the step edge rows consist of
Si atoms or dangling bonds rather than Ag-related features, since
on the terraces Ag atoms preferentially adsorb on Si-related
sites.

In summary, we have shown that Ag adsorption on vicinal Si(111)
can be used for fabrication of ordered nanoscale arrays. The Ag
induced \iiixi\ reconstruction stabilizes a well ordered step
array on vicinal Si(111) miscut towards [11$\bar{2}$], preventing
the equillibrium step bunching characteristic of the clean surface
\cite{phaneuf_88a}. The main terrace width changes by multiples of
the \iiixi\ unit cell size, 3.8 nm being the preferred width for
6$^\circ$ miscut samples. By matching the macroscopic miscut to
the \iiixi\ cell size, the sample homogeneity on the $\mu$m scale
may be further improved. Filled and empty states STM images of the
detailed terrace structure are in accord with the previously
proposed HCC model \cite{erwin_98a}. Using the \iiixi\
reconstructed step array as a template, one-dimensional chains of
Ag nanodots can be grown, displaying a preferential nearest
neighbor separation of four lattice constants along [1$\bar{1}0$].
Such dot arrays could eventually be used for future atomic scale
memories, as recently proposed for Si(111)-(5$\times$2)-Au
\cite{bennewitz_02a}.

{\bf Acknowledgment} J. K. is supported by the Marie Curie grant
HPMF-CT-1999-00280. A.M. and J.E.O. are supported by the
Universidad del Pa\'{\i}s Vasco
(1/UPV/EHU/00057.240-EA-8078/2000).



\begin{references}
\bibitem{leedremark}
The terrace width given here is derived from the maxima of beam intensities at a few energies and is not corrected for the
spot profile. This could slightly overestimate the average terrace size, depending on the shape of the terrace
 width distribution \protect\cite{wollschlaeger_97a}.
\end{references}
\end{document}